\documentclass[a4paper]{article} 

\def\comment#1{} 
\def\journalfont{\rm}         
\def\jou#1{{\journalfont #1\ }}
\def\joudef#1#2{\def #1{\jou{\ignorespaces #2}}}

\joudef{\aaa}    { Astron.\ Astrophys.}
\joudef{\aip}    { Adv.\ Phys.}
\joudef{\adm}    { adv.\ math.}
\joudef{\am}     { Ann.\ Math.}
\joudef{\apl}    { Ann.\ Phys.\ (Leipzig)}
\joudef{\apny}   { Ann.\ Phys.\ (N.Y.)}
\joudef{\arnps}  { Annu.\ Rev.\ Nucl.\ Part.\ Sci.}
\joudef{\apj}    { Astrophys.\ J.}
\joudef{\apjl}    { Astrophys.\ J.\ Lett.}
\joudef{\cjp}    { Can.\ J.\ Phys.}
\joudef{\cmp}    { Commun.\ Math.\ Phys.}
\joudef{\cqg}    { Class.\ Quantum Grav.}
\joudef{\grg}    { Gen.\ Rel.\ Grav.}
\joudef{\ijmpd}  { Int.\ J.\ Mod.\ Phys.\ D}
\joudef{\ijtp}   { Int.\ J.\ Theor.\ Phys.}
\joudef{\invm}   { Invent.\ Math.}
\joudef{\jm}     { J.\ Math.}
\joudef{\jmaa}   { J.\ Math.\ Anal.\ Appl.}
\joudef{\jmp}    { J.\ Math.\ Phys.}
\joudef{\jpa}    { J.\ Phys.\ A}
\joudef{\lr}    { Liv.\ Rev.\ Rel.}
\joudef{\mnras}  { Mon.\ Not.\ R.\ Ast.\ Soc.}
\joudef{\mpl}   { Mod.\ Phys.\ Lett.} 
\joudef{\mpla}   { Mod.\ Phys.\ Lett.\ A} 
\joudef{\nature} { Nature}
\joudef{\nc}     { Nuovo Cim.}
\joudef{\npb}    { Nuc.\ Phys.\ B}
\joudef{\ph}     { Physica}
\joudef{\pla}    { Phys.\ Lett. A}
\joudef{\plb}    { Phys.\ Lett. B}
\joudef{\pr}     { Phys.\ Rev.}
\joudef{\pra}    { Phys.\ Rev.\ A}
\joudef{\prb}    { Phys.\ Rev.\ B}
\joudef{\prc}    { Phys.\ Rev.\ C}
\joudef{\prd}    { Phys.\ Rev.\ D}
\joudef{\prep}   { Phys.\ Rep.}
\joudef{\prl}    { Phys.\ Rev.\ Lett.}
\joudef{\prsla}  { Proc.\ Roy.\ Soc.\ Lond.\ A}
\joudef{\ptp}    { Prog.\ Theor.\ Phys.}
\joudef{\ptps}   { Prog.\ Theor.\ Phys.\ Suppl.}
\joudef\rmp      { Rev.\ Mod.\ Phys.}
\joudef\spj      { Sov.\ Phys.\ JETP}

\catcode`@=11

\def\eqalign#1{\null\,\vcenter{\openup\jot\m@th
  \ialign{\strut\hfil$\displaystyle{##}$&$\displaystyle{{}##}$\hfil
      \crcr#1\crcr}}\,}
\def\meqalign#1{\null\,\vcenter{\openup\jot\m@th
  \ialign{\strut\hfil$\displaystyle{##}$&&$\displaystyle{{}##}$\hfil
      \crcr#1\crcr}}\,}
\catcode`@=12   

\newdimen\arrayruleHwidth
\makeatletter
\def\Hline{\noalign{\ifnum0=`}\fi\hrule \@height \arrayruleHwidth
  \futurelet \@tempa\@xhline}
\makeatother

\newcommand\thickbaselines{\baselineskip=20pt\lineskip=3pt\lineskiplimit=3pt}
\catcode`@=11
\def\cases#1{\left\{\,\vcenter{\thicknormalbaselines\m@th
             \ialign{$##\hfil$&\quad##\hfil\crcr#1\crcr}}\right.}
\def\matrix#1{\null\,\vcenter{\thickbaselines\m@th
    \ialign{\hfil$##$\hfil&&\quad\hfil$##$\hfil\crcr
      \mathstrut\crcr\noalign{\kern-\baselineskip}
      #1\crcr\mathstrut\crcr\noalign{\kern-\baselineskip}}}\,} 
\catcode`@=12   

\newcommand{\eprint}{\textsf} 

\newcommand\be{\begin{equation}} \newcommand\ee{\end{equation}} 
\newcommand\bd{\begin{displaymath}}\newcommand\ed{\end{displaymath}}

\renewcommand{\d}{{\rm d}} 

\newcommand\ts\textstyle

\def\undersim#1{\mathop{\vtop{\ialign{##\crcr
     $\hfil\displaystyle{#1}\hfil$\crcr\noalign
     {\kern1pt\nointerlineskip}\hbox{$\hfil\sim\hfil$}\crcr
     \noalign{\kern1pt}}}}}

\newcommand{\acronym}[3]{\newcommand{#1}{#3 (#2)\relax\renewcommand{#1}{#2}}}

\def\eg{{\it e.g.}} \def\etal{{\it et al.}} \def\ie{{\it i.e.}}

\newtoks\reportnoregister \newtoks\eprintnoregister
\newcommand{\reportnumber}[1]{\reportnoregister={#1}}
\newcommand{\eprintnumber}[1]{\eprintnoregister={#1}}

\reportnumber{\mbox{}} 
\eprintnumber{\mbox{}} 

\newcommand{\reportid}{
   \begin{minipage}{17cm}\vspace{-3.2cm}
     \begin{flushright}
      {\normalsize \the\reportnoregister \\[-.2cm]
            \eprint{\the\eprintnoregister}}\vspace{3.2cm}
     \end{flushright}
   \end{minipage}\hspace{-17cm} }

\catcode`@=11   
\def\title#1{\gdef\@title{\reportid#1}}
\catcode`@=12   

\acronym{\SSS}{SSS}{{\em static spherically symmetric}}
\acronym{\TOV}{TOV}{{\em Tolman-Oppenheimer-Volkoff}}

\newcommand{\ncd}{\newcommand}
\ncd{\nms}{\negmedspace}
\ncd{\nts}{\negthickspace}
\ncd{\mcl}[1]{\mathcal{#1}}
\ncd{\beq} {\begin{equation}}
\ncd{\eeq} {\end{equation}}
\ncd{\BE} {\begin{eqnarray}}
\ncd{\EE} {\end{eqnarray}}
\ncd{\rarr} {\rightarrow}
\ncd{\larr} {\leftarrow}
\ncd{\lrarr} {\leftrightarrow}
\ncd{\lbeq}[1]  {\label{eq: #1}}
\ncd{\refeq}[1] {(\ref{eq: #1})}
\ncd{\mrm}    {\mathrm}
\ncd{\nn}{\nonumber}
\ncd{\mbf}[1] {{\mathbf #1}}
\ncd\T{\frac{1}{2}h^{\mu\nu}p_\mu p_\nu}
\ncd{\ms}{\mathstyle}
\ncd{\ds}{\displaystyle}
\ncd{\Yc}{Y_{\rm c}}
\ncd{\Yorb}{Y_{\rm orb}}

\ncd{\der}{{\mathrm{d}}}
\ncd{\rtil}{\tilde{r}}
\ncd{\Mr}{\frac{2M}{r}}
\ncd{\rhotil}{\tilde{\rho}}
\ncd{\rstar}{r_{*}}

\usepackage{amsmath}
\usepackage{graphicx}
\usepackage{amssymb}

\ncd{\tsfrac}[2]{{\ts\frac{#1}{#2}}}
\ncd{\D}{\mathcal{D}}
\ncd{\Man}{\mathcal{N}}
\ncd{\E}[1]{\times 10^{#1}}

\begin{document}

\title{A characteristic approach to the quasi-normal mode problem}
\author{Lars Samuelsson$^1$\footnote{E-mail: \eprint{lars@soton.ac.uk}}, 
        Nils Andersson$^1$\footnote{E-mail: \eprint{na@maths.soton.ac.uk}}
        ~and Asimina Maniopoulou$^1$ \\[-10pt]
{\small $^1$ School of Mathematics, University of Southampton, Southampton SO17~1BJ, UK} \\
\begin{minipage}[t]{0.8\linewidth}\small{ 
\abstract{In this paper we discuss a new approach to the quasi-normal mode 
problem in general relativity. By combining a characteristic
formulation of the perturbation equations with the integration of a
suitable phase-function for a complex valued radial coordinate, we
reformulate the standard outgoing-wave boundary condition as a zero
Dirichlet condition. This has a number of important advantages over
previous strategies. The characteristic formulation permits coordinate
compactification, which means that we can impose the boundary
condition at future null infinity. The phase function avoids
oscillatory behaviour in the solution, and the use of a complex radial
variable allows a clean distinction between out- and ingoing waves. We
demonstrate that the method is easy to implement, and that it leads to
high precision numerical results. Finally, we argue that the method
should generalise to the important problem of rapidly rotating neutron
star spacetimes. }}
\end{minipage}}

\date{}

\maketitle

\section{Introduction}

The dynamical oscillations of compact objects is a problem of great
relevance for general relativistic astrophysics. Interesting
questions range from observational to theoretical, with the
possibility of detecting gravitational waves from pulsating neutron
stars and black holes and using the signals to infer source parameters
providing strong motivation for detailed studies. From the more
theoretical point of view, the stability properties of these
astrophysical bodies are intimately linked to their oscillation
spectra. For rotating neutron stars, this is clearly demonstrated by
the gravitational-wave driven instability that was first discussed by
Chandrasekhar \cite{chandra:cfs}, Friedman and Schutz \cite{fs:cfs}.

In general relativity, non-radial fluid oscillations radiate
gravitational waves. Hence the oscillation mode problem is
conceptually different from that in Newtonian gravity. In order to
determine the pulsation modes one must impose an outgoing-wave
boundary condition at infinity.  It is well documented that this leads
to technical difficulties if the “quasi-normal” modes are rapidly
damped. The main reason for this is that on a spacelike hypersurface,
as obtained by assuming that the perturbation quantities have a
$\exp(i\omega t)$ time-dependence, the required solutions
grow exponentially towards infinity. In order to impose an accurate
outgoing-wave condition one must be able to filter out the
exponentially decaying ingoing-wave component.

Various methods have been devised to handle this problem, see the
review by Kokkotas and Schmidt \cite{ks:qnm} for an exhaustive
discussion. One approach, that has the advantage that it is relatively
easy to implement, is based on the use of analytic continuation and
integration of the perturbation equations for complex values of the
radial variable. The method, which was first used to calculate
accurate quasi-normal modes of black holes \cite{andersson:bhqnm}, has
been successfully applied to the stellar oscillation problem \cite{aks:qnm}.

The main outstanding problem in this area concerns the oscillations of
rapidly spinning relativistic stars. In this case the perturbation
equations that need to be solved in the exterior vacuum are no longer
describable by a single wave-type equation (essentially because the
spacetime is no longer of Petrov type D \cite{bwmb:rot}). This makes
the solution more involved and, in particular, any method that relies
on the solution of a single separated differential equation no longer
applies. This problem has yet to be overcome. The best results that we
have correspond on the one hand to the so-called neutral modes
\cite{sf:neutral,msb:neutral}, identifying the points where the
fundamental f-mode becomes susceptible to the
Chandrasekhar-Friedman-Schutz instability, and on the other hand to
modes determined after ignoring the metric perturbations (within the
relativistic Cowling approximation) \cite{ye:rotcowling}. In the latter
case one can use multipole formulae to estimate the damping/growth
rate of the modes. Nevertheless, the situation is not
satisfactory. After all, it is not clear how accurate the Cowling
approximation will be for the various classes of modes that one may be
interested in. There is a clear need for a more detailed solution to
the mode-problem for rapidly rotating neutron stars.

This paper introduces a new strategy to the problem. The main idea is
to use a characteristic formulation of the perturbation problem
\cite{mina:thesis} to avoid the diverging eigenfunctions that plague
the standard analysis. By considering two simple model problems, we
will demonstrate the promise of this approach. Most importantly, we
will show how one can formulate the outgoing-wave condition as a
Dirichlet condition for a suitably introduced phase-function
\cite{ga:quickndirty}. Taken together with the fact that a
characteristic formulation permits coordinate compactification without
the loss of resolution \cite{ccm:V}, this enables us to impose the
desired boundary condition at future null infinity with extreme
precision. We illustrate the key features of the new method by
applying it to the well-studied problem of axial spacetime modes of a
uniform density star. Using a spectral approach to solve for the
relevant phase-function in the exterior of the star, we obtain results
with very high numerical precision. The reliability of the method is
shown by considering an ultra-compact star, with $R/M=2.26$, which has
both very slowly and very rapidly damped modes. We confirm previous
results for the slowly damped modes, and even provide some new results
by identifying several previously unknown interface $w$-modes. This
result is likely irrelevant from the astrophysics point of view, but
it is conceptually interesting since it hints at the possibility that
there may exist an infinite number of such modes.

The numerical results that we report are in themselves not
particularly interesting. The main achievement is the successful new
formulation of the quasi-normal mode problem that should be 
“straightforward” to extend to the rapid rotation case. The
characteristic formulation of the Einstein equations is of course
well-known, and we cannot see any real difficulties in generalising
our method to the resultant coupled perturbation equations. In fact,
the possibility of having a Dirichlet condition at infinity in the
computationally more involved two-dimensional problem is very
attractive. Some details obviously remain to be worked out. The main
issue concerns phase-functions for coupled equations, a problem that
has already been considered in quantum scattering problems
\cite{kalle:phase}. Hence, we feel very optimistic. For the first time
we have a truly promising strategy for dealing with one of the most
challenging problems in this area of research, the determination of
oscillation modes and the associated gravitational-wave damping/growth
rate for fast spinning stars.

\subsection{Problem setup}

As a model problem we will consider the mode problem in static
spherically symmetric spacetimes. In particular, the wave equations we
will discuss in this paper can all be written on the form
\beq\lbeq{genwave}
  (\D_a\D^a-U)\psi = 0
\eeq
where $\D_a$ is the covariant derivative on the submanifold orthogonal
to the spherical symmetry surfaces ($a$ runs from~0 to 1) and $U$ is
an effective potential depending only on the radial coordinate. For
vacuum perturbations the potential is of the form
\beq \lbeq{Utot}
  U = \frac{l(l+1)}{r^2} - \frac{6M}{r^3}\mathcal{U}_{\pm}
\eeq
where 
\begin{align}
  \mathcal{U}_{-} &= 1 \lbeq{Uax}\\
  \mathcal{U}_{+} &= \frac{[l^2(l+1)^2 - 4]r^2 + 12M(r-M)}{[(l-1)(l+2)r + 6M]^2}\lbeq{Upol}
\end{align}
for axial and polar perturbations, respectively. Here $M$ is the mass,
$r$ is the Schwarzschild radial coordinate and $l\ge2$ is the usual
angular ``quantum number'' originating from the separation of the
angular dependence. Asymptotically the polar piece approaches a
constant;
\beq
  \mathcal{U}_{+} \sim \frac{l(l+1)+2}{l(l+1)-2}
\eeq
and it is easy to show that $\frac{5}{7}\le \mathcal{U}_{+} \le 2$
everywhere outside $r=2M$. It is evident that the asymptotic behaviour
of the axial and polar perturbations is very similar. For this reason
we restrict our attention, from now on, to the axial sector.

We shall later, as a test case, consider axial perturbations of
perfect fluid stars. These are also governed by an equations of the
form
\refeq{genwave}, with $U$ given by
\beq\lbeq{Upf}
 U = \frac{l(l+1)}{r^2} - \frac{6m}{r^3} + \frac{\kappa}2(\rho-p)
\eeq 
where $m=m(r)$ is the mass inside $r$, $\rho$ is the energy density
and $p$ is the pressure \cite{cf:osc,kokkotas:axialmodes}.

In our analysis of the wave equation \refeq{genwave} we shall use our
freedom of choice of coordinates. It is conventional to adapt the
coordinates to the timelike Killing vector on the background and
choose the remaining radial coordinate to be either the Schwarzschild
area radial function $r$ or the Regge-Wheeler tortoise radius
$\rstar$. The former has the advantage of being invariantly defined as
giving the area of a spherical symmetry surface through Area~$=4\pi
r^2$. On the other hand, the tortoise coordinate is intimately related
to the characteristics of the wave equation for massless fields and is
related to $r$ (in vacuum) through
\beq\lbeq{rstar}
  \rstar = r + 2M\ln[C(r-2M)]
\eeq
where $C$ is a constant related to translations.  Since we are
primarily interested in the asymptotics (which does not depend
strongly on whether $r$ or $\rstar$ is used) we shall use $r$ in the
following.

An alternative to using the Killing time as a coordinate is to use 
\emph{characteristic} coordinates as introduced by Bondi \etal 
\cite{bbm:characteristic} and Sachs \cite{sachs:characteristic}.
The relation between the different coordinates is
\begin{align}
  x &= r \\
  u &= t-\rstar(r) 
\end{align}
where $u$ is the retarded time and the characteristic radial variable
$x$ is actually the same as the area radius and we shall soon
denote it by $r$. However it is important to remember that
\beq\lbeq{dr}
  \frac{\partial}{\partial r} = \frac{\partial x}{\partial r}\frac{\partial}{\partial x} 
     + \frac{\partial u}{\partial r}\frac{\partial}{\partial u} = 
   \frac{\partial}{\partial x} - \frac{\d \rstar}{\d r}\frac{\partial}{\partial u}
\eeq

For vacuum spacetimes and for the coordinates discussed above the line
elements can be written
\beq
  \d s^2 = \d s_j^2 + r^2(\d\theta^2 + \sin^2\theta\d\phi^2)
\eeq
where 
\begin{align}
  \d s_j^2 &= -e^{2\nu}\d u^2 - 2\d u\d r \qquad \mbox{Bondi-Sachs (B)} \\
  \d s_j^2 &= -e^{2\nu}\d t^2 + e^{-2\nu}\d r^2 \qquad \mbox{Schwarzschild (S)}
\end{align}
We have set $x=r$ and introduced the notation
\beq \lbeq{enu}
  e^{2\nu} = 1-\frac{2M}{r}
\eeq
The wave equation \refeq{genwave} can now be explicitly written down
\begin{align}
  -2\psi_{,ur} + e^{2\nu}\psi_{,rr} + \frac{2M}{r^2}\psi_{,r} - U\psi=0 \qquad &\mbox{(B)} \lbeq{wbs} \\
  -e^{-2\nu}\psi_{,tt} + e^{2\nu}\psi_{,rr} + \frac{2M}{r^2}\psi_{,r} - U\psi=0 \qquad &\mbox{(S)} \lbeq{ws}
\end{align}
Note that, in the characteristic formulation, there is no second
derivative with respect to the ``time'' $u$. To find mode solutions we
take the time dependence to be given by $\psi\propto
e^{i\omega T}$ where $T=t,u$ depending on coordinates. We let primes
denote derivatives with respect to the spatial coordinate (remember
that the primes are different due to
\refeq{dr}!).
\begin{align}
  e^{2\nu}\psi'' + 2\left(\frac{M}{r^2} - i\omega\right)\psi' -U\psi = 0 \qquad &\mbox{(B)}\\
  e^{2\nu}\psi'' + \frac{2M}{r^2}\psi' - (U - e^{-2\nu}\omega^2)\psi=0 \qquad &\mbox{(S)}
\end{align}
We may note here that the algebraic transformation $\psi \rightarrow
e^{-i\omega\rstar}\psi$ brings the Schwarzschild equation to the form
of the Bondi-Sachs equation. Thus for the simple case of static
spacetimes considered here we can mimic the characteristic approach by
this simple transformation. However, the case we are really interested
in is the two-dimensional rapidly rotating problem where such a trick
is not likely to exist.

\section{A pedagogical toy problem}

\begin{figure}
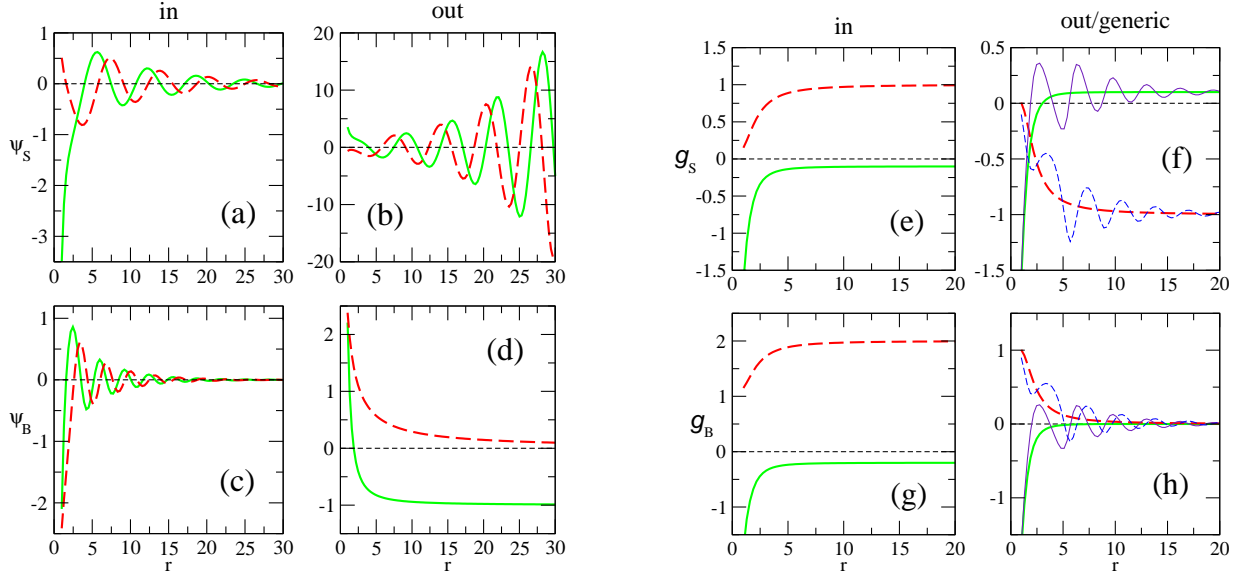

  \includegraphics[width=0.45\textwidth, clip]{psi.eps}\phantom{MMMM}
  \includegraphics[width=0.45\textwidth, clip]{g.eps} \caption{The
  left set of panels (a-d) display the real (solid) and imaginary
  (dashed) parts of the wave function $\psi$ as a function of $r$. The
  right set (e-h) shows the phase function $g(r)$. In each set of four
  panels the top two (a-b and e-f) show the solution in Schwarzschild
  coordinates whereas the bottom two (c-d, g-h) display the
  characteristic solutions. The left panels (a,c and e,g) show the
  ingoing solutions while the right panels (b,d and f,h) instead
  display the outgoing solution. In panels f and h we additionally
  plot a generic solution with $C_{\mrm{in}}=0.57$,
  $C_{\mrm{out}}=1.3$. See the main text for further
  discussion.}\label{fig:r}
\end{figure}

Since the main interest in this paper is the behaviour of wave
solutions in the asymptotically flat region of spacetime far away from
the source we will begin by considering the very illustrative toy
problem in which $M$ is set to zero. It should be intuitively clear,
and it will later be demonstrated, that this problem has the same
properties near infinity as the $M\neq0$ case. A great advantage is
that it allows for an exact solution. In fact, in Schwarzschild
coordinates
\beq
  \psi = \sqrt{r}[C_{\mrm{in}}H^{(1)}_{l+1/2}(\omega r) 
    + C_{\mrm{out}}H^{(2)}_{l+1/2}(\omega r)] \qquad \mbox{(S)}
\eeq
where $H^{(1)}$ and $H^{(2)}$ represent the in- and outgoing Hankel
(or Bessel of the third kind) functions and the constants refer to the
proportion of out- and ingoing waves as can be seen from the
asymptotic behaviour
\beq
  \psi e^{i\omega t} \sim C_{\mrm{in}}e^{i\omega(t+r)} 
    + C_{\mrm{out}}e^{i\omega (t-r)} \qquad \mbox{(S)}
\eeq
As noted previously, in Bondi-Sachs coordinates the solution is just
given by multiplying the Schwarzschild solution by
$e^{i\omega\rstar}=e^{i\omega r}$.  In order to illustrate the nature of
these solutions we will now examine the special case of $l=2$. Then
the solution becomes, after some redefinitions of the constants,
\beq
   \psi = C_{\mrm{in}}\frac{\omega^2r^2 + 3i\omega r - 3}{\omega^2r^2}e^{i\omega r} 
 - C_{\mrm{out}}\frac{\omega^2r^2 - 3i\omega r - 3}{\omega^2r^2}e^{-i\omega r}  \qquad \mbox{(S)}
\eeq
For definiteness we will plot the solutions for an arbitrarily chosen
frequency $\omega=1+i/10$ in units such that the inner boundary is
located at $r=1$. The ingoing and outgoing parts of the solution in
Schwarzschild coordinates are plotted in figures
\ref{fig:r}a and \ref{fig:r}b, respectively. If we imagine finding
these solutions numerically we immediately note two problems. First
the solutions are oscillating. This is clearly a problem since we are
really interested in imposing boundary conditions at infinity and
hence would like to compactify spacetime in order to cover it on a
finite grid. It is evident, regardless of the number of grid points,
that the oscillating behaviour will not be resolved at some finite
radius.  The other problem is that the ingoing solution is
exponentially decreasing outwards. Hence, it will be numerically
impossible to distinguish the outgoing solution from a solution
contaminated by an ingoing piece.  In figures \ref{fig:r}c and
\ref{fig:r}d we instead plot the characteristic/Bondi-Sachs form. The
outgoing solution is clearly much better behaved here and would in
principle be possible to track numerically on a compactified
grid. However, the problem with the rapid decay of the ingoing
solution remains.

The oscillation problem is often resolved by introducing a phase
function. Several options exist which are essentially equivalent, see
\eg\ \cite{ga:quickndirty}. Here we simply define 
the phase function
to be
\beq\lbeq{phaseg}
  g = \frac{\psi'}{\psi}
\eeq
as suggested from WKB-type arguments.  
In figure \ref{fig:r}e-h we show the behaviour of the phase
function. Panels \ref{fig:r}e and \ref{fig:r}f show the ingoing and
outgoing solutions in Schwarzschild coordinates whereas \ref{fig:r}g
and \ref{fig:r}h are the Bondi-Sachs forms. In panels \ref{fig:r}f and
\ref{fig:r}h we additionally show a generic solution with coefficients
arbitrarily chosen to be $C_{\mrm{in}}=0.57$ and
$C_{\mrm{out}}=1.3$. We see that although the phase function solves
the problem of oscillating solutions we still cannot distinguish the
outgoing solution from a generic one at infinity. The reason is that
the transformation to the variable $g$ is non-linear and pieces coming
from the ingoing solution will always be exponentially suppressed for
positive imaginary parts of $\omega$. It is clear that we must find a
way to remove the exponential decay of the ingoing solution in order
to find a successful numerical scheme. Before turning to this problem
it is useful to note that although all solutions are asymptotically
constant, the outgoing Bondi-Sachs version has the advantage of being
asymptotic to zero. Hence, in that formulation, the outgoing wave
boundary condition can be replaced by a much simpler zero Dirichlet
condition.

\begin{figure}
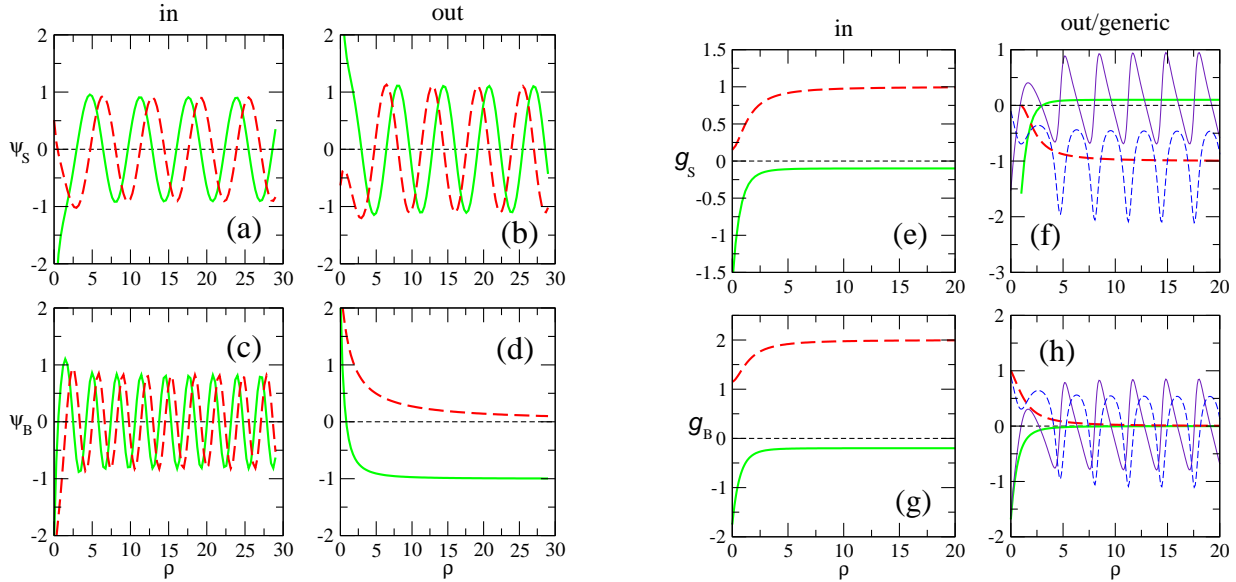

  \includegraphics[width=0.45\textwidth,
  clip]{psirho.eps}\phantom{MMMM}
  \includegraphics[width=0.45\textwidth, clip]{grho.eps} \caption{Same
  as figure \ref{fig:r} but with $\cpr$ as radial coordinate. Hence, these solutions are
obtained from integration in the complex $r$-plane.}\label{fig:rho}
\end{figure}

We turn now to the decay of the ingoing solution.  The problem comes
from the asymptotic factor
\beq\lbeq{exp}
  e^{i\omega r} = e^{i\Re(\omega)r}e^{-\Im(\omega)r}
\eeq
There is a neat trick to resolve this problem which has been used for
quasi-normal modes, see \eg\ \cite{andersson:bhqnm,aks:qnm}. The idea is to
analytically continue the support of the dependent variable $g$ or
$\psi$ and let the radial coordinate take on complex values. In this
spirit we write 
\ncd{\cpr}{\varrho}
\beq\lbeq{rcplx}
  r = R + \cpr e^{i\vartheta}
\eeq
where $R$ is the (real) radius outside which we are interested in the
solution and $\cpr$ and $\vartheta$ are real. We now wish to choose the
angle $\vartheta$ in the complex plane such that the decay of the ingoing
solution is avoided. Substituting \refeq{rcplx} into \refeq{exp} we see
that we must choose 
\beq
  \vartheta = -\arg(\omega)
\eeq
Having made this choice\footnote{Obviously, $\omega=0$ is a degenerate
case in this respect and cannot be treated with the present
approach. However, since the $\omega=0$ problem for rotating stars has already been
studied in detail by Stergioulas and coworkers
\cite{sf:neutral,msb:neutral} this is not a serious drawback of our
method.} we plot the various solutions again in figure
\ref{fig:rho}. Note that all divergent/decaying behaviour is gone. 
We see that, in principle, there exist three possibilities to
implement a successful numerical scheme. The characteristic approach
aided by the complex radial coordinate as displayed in panels
\ref{fig:rho}c-d allows distinction of the in- and outgoing solutions 
together with compactification and imposition of boundary conditions
at or near infinity. The same applies to the phase function approaches
as shown in figures \ref{fig:rho}e-f (Schwarzschild) and
\ref{fig:rho}g-h (Bondi-Sachs). However, the Bondi-Sachs phase function 
has the advantage of going asymptotically to zero. Hence, in this
approach, the outgoing-wave boundary condition may be replaced by a
zero Dirichlet condition on the phase function at null infinity. This
is of course a great simplification when extensions to more general
circumstances (such as rapidly rotating sources) are considered.  

\subsection{A numerical example}

In order to show that the approach can be accurately implemented we
now solve a toy quasi-normal mode problem. As above we take 
$M=0$, and choose units in which $R=1$. To complete the problem we
choose boundary conditions such that, in Bondi-Sachs coordinates 
\beq
  \psi'|_{r=R} = g|_{r=R} = 0
\eeq
whereas, as shown above, $g$ satisfies a zero Dirichlet condition at
infinity. The inner boundary condition is chosen purely for
mathematical convenience. In order to be able to enforce the boundary
conditions at infinity we compactify spacetime by introducing the
radial coordinate $x$ through
\beq
  x = \frac{r_1-\cpr}{r_1+\cpr}
\eeq
where the real constant $r_1$ makes it easier to keep track of the
dimensions. Note that 
\begin{align}
  r = R      \quad &\leftrightarrow \quad \cpr=0      \quad \leftrightarrow \quad x=1 \\
  r = \infty \quad &\leftrightarrow \quad \cpr=\infty \quad \leftrightarrow \quad x=-1 
\end{align}
so that the exterior spacetime is mapped to the domain $-1< x \le
1$. The reason we chose this range is that we base our numerical code
on a Chebychev pseudo-spectral method, \eg\
\cite{boyd:spectral}. Using equations \refeq{wbs} and \refeq{phaseg}
we find that the phase function has to satisfy
\beq\lbeq{g0}
  g_{,x} = \frac{2r_2}{(x+1)^2}g^2 - \frac{4i\omega r_2}{(x+1)^2}g - \frac{2r_2 l(l+1)}{[R(1+x) + r_2(1-x)]^2}
\eeq
where $r_2=r_1e^{i\vartheta}$. As will be seen in the next section,
the irregular singular point at infinity ($x=-1$) does not pose a
problem for the outgoing solution since $g\sim(1+x)^2$ in the
neighbourhood of null infinity. The purely ingoing solution is (of
course) also well behaved since it asymptotically cancels the two
first terms on the right hand side of \refeq{g0}. When we enforce the
boundary conditions at infinity we add the point $x=-1$ and put $g=0$
exactly there. This is one of the advantages of the pseudo-spectral
approach -- we are actually able to enforce the boundary conditions
\emph{at} infinity. The family of solutions to \refeq{g0} then provide
a function $g_s(\omega) = g(x=1,
\omega)$ whose zeros correspond to the quasi-normal modes. 
At a practical level, we solve equation \refeq{g0} by a Chebychev
pseudo-spectral code using a Newton-Kantorovich scheme to turn the
problem from a non-linear equation to iteratively solved linear
equations (see \eg\ the book by Boyd \cite{boyd:spectral} for details
on spectral methods). We iterate until the corrections to the solution
are everywhere less than $10^{-15}$. We solve the equation for a large
set of frequencies. A contour plot of the absolute value of
$g_s(\omega)$ is then used to locate the points in the complex
frequency plane where $g_s(\omega)$ is close to zero. These points are
then used as initial guesses in a M\"uller type root finder. We
decided to use a spectral code since we expect this approach to
generalise to the coupled partial differential equations which need to
be solved in the rapidly rotating case.

\begin{table}
\begin{center}
\begin{tabular}{|l|l|l|}
 \hline
  $l$ & Frequencies, exact & numerical \\
 \hline
  2 & $2i$                          & $0.000000000 + 2.000000000i$ \\
  3 & $\frac12(\sqrt{5} + 5i)$      & $1.118033989 + 2.500000000i$ \\
  7 & $5.071731535 + 3.511061460i$  & $5.071731535 + 3.511061460i$ \\ 
    & $2.947010505 + 4.736313493i$  & $2.947+ 4.74i$ \\ 
    & $0.9710593605 + 5.252625048i$ & $0.97106 + 5.2526i$ \\
 \hline                               
\end{tabular}
\caption{Quasi-normal modes of the toy problem for various $l$. All the digits 
shown have converged and agree with the exact
solution. In some cases we reach machine precision, see table
\ref{tab:convergence}. Note, however, the poor convergence for some of the 
$l=7$ modes. See the main text for an explanation of this.}\label{tab:f}
\end{center}
\end{table}

We examine the cases $l\in\{2,3,7\}$. The eigenmodes in the
positive quadrant are given in table \ref{tab:f}.
For the case $l=2$ we see that we have the extreme case of a ``mode''
with purely imaginary frequency. In table \ref{tab:convergence} we
show the convergence of the solution as a function of grid points
$N$. As we can see machine precision (in this case double $\sim
10^{-16}$) is reached exponentially fast, a trade mark of a spectral
code. The convergence  is equally impressive for  $l=3$. For the
case $l=7$, however, we run into problems. The first mode ($\omega
\approx 5.07 + 3.51i$) is found without problems, but as the argument
of the frequency starts growing the code runs into convergence
problems. The reason for this behaviour can be seen from the exact
solution. For any $l\ge2$ the solutions have the general form (in
terms of $r$)
\beq
  g = \frac{\sum_{i=0}^{l-1}a_i(l) (\omega r)^i}{r\sum_{i=0}^{l}b_i(l) (\omega r)^i}
\eeq
where $a_i(l)$ and $b_i(l)$ are constants depending on $l$.  As we
can see, these functions have a number of poles determined by the
zeros of the polynomial in the denominator. If any of these poles
happen to lie close to our path of integration we can expect large
variations in $g$. This is indeed what happens for $l=7$, causing
convergence problems for the spectral code (which relies crucially on
the smoothness of the solution). We have tried to use other 
integration routines and found it straightforward to improve the
accuracy of the obtained frequencies. However, the main goal here is
not to obtain accuracy in this highly unphysical toy problem, but
rather to assess the pros and cons of our proposed characteristic scheme. Hence, the
convergence problems displayed in table \ref{tab:f} only serve as a
nice reminder of the kind of problems that can be encountered in real
situations. We should note here that if a pole should happen to lie
exactly in the path of integration the method presented here will
fail. This seems highly unlikely to happen however and is not related
to the characteristic approach, but rather to the analytic
continuation and introduction of a complex $r$.
\begin{table}
\begin{center}
\begin{tabular}{|r|r|r|r|}
 \hline
  $N$  &  Frequency & $2-|\omega|$ & $\frac12\pi-\arg(\omega)$ \\
 \hline
  $40$  & $5.444541\E{-3}   + 1.997613i$ & $-2.379953\E{-3}$  & $-2.725517\E{-3}$ \\
  $80$  & $4.063484\E{-7}   + 2.000000i$ & $4.089420\E{-8}$   & $-2.031742\E{-7}$ \\
  $120$ & $1.893739\E{-11}  + 2.000000i$ & $-1.279776\E{-11}$ & $-9.468648\E{-12}$ \\
  $160$ & $3.011046\E{-15}  + 2.000000i$ & $1.332268\E{-15}$  & $-1.554312\E{-15}$ \\
  $200$ & $-4.551599\E{-16} + 2.000000i$ & $4.440892\E{-16}$  & $2.220446\E{-16}$ \\
  $240$ & $9.935105\E{-16}  + 2.000000i$ & $-6.661338\E{-16}$ & $-4.440892\E{-16}$\\
 \hline
\end{tabular}
\caption{Convergence test for the $l=2$ quasi-normal mode of the toy problem. The 
two right-hand columns show the error in the numerically obtained absolute
value and phase of the frequency as compared with the exact value
$\omega=2i$. Machine precision is reached exponentially fast as the resolution ($N$)
is increased.} \label{tab:convergence}
\end{center}
\end{table}

\section{Quasi-normal modes of uniform density ``stars''.}

We now turn to the slightly more relevant problem of finding the
axial $w$-modes of a constant density star. This problem has been
treated many times before (\eg\
\cite{cf:gwresonance,kokkotas:axialmodes,akk:oscillation,tsm:gw}) 
allowing us to compare our results to the existing ones. Since the
behaviour near null infinity is most strongly dependent on the
frequency and not on the particulars of the wave function(s) near or
in the star we believe that this comparatively simple problem serves
as a sufficient test of our method as far as non-rotating stars are
concerned. In order to span as large a region of frequency space as
possible we concentrate on an ultra compact model for which $R=2.26M$,
\ie\ quite close to the Buchdahl limit $R=2.25M$. The $w$-mode
spectrum \cite{ks:w-modes} can then be loosely divided into three
parts (see
\cite{ks:qnm}). The
\emph{trapped modes} \cite{cf:gwresonance} are characterised by a low damping rate, \ie\ a
small imaginary part of the frequency. These modes exist due to a peak
in the potential near $r=3M$ and correspond, in a loose sense, to
almost bound states of the Schr\"odinger-like wave equation. For higher
``energies'' the modes become less and less bound (\ie\ more damped),
having increasingly large imaginary parts of the frequencies. These
are the ordinary, or curvature, $w$-modes. An infinite number of these modes
is likely to exist.

There also exist another branch of modes, the \emph{interface}, or
$w_{I\!I}$-modes \cite{lns:w2}. These are thought to correspond to scattering of
gravitational waves off the ``surface''  of the star
much like hard-sphere scattering of sound waves. These rapidly damped
modes have large imaginary parts of the frequency. Previous numerical
surveys have only found a few (typically two or three) such
modes. With our method we are able to probe larger imaginary parts and
discover many more $w_{I\!I}$-modes, thus raising the question whether
these modes also form an infinite family.

We now proceed to show that the analysis carried out in the preceding
section generalises to spherically symmetric spacetimes with arbitrary
$M$. We shall only consider the characteristic approach here. It is
straightforward to show that the wave function has the asymptotic
behaviour 
\begin{align}
  \psi_{\mrm{out}} &\sim 1 - \frac{il(l+1)}{2\omega r} + \frac{12iM\omega + 2l(l+1) + l^2(l+1)^2}{8\omega^2r^2} + \ldots \\
  \psi_{\mrm{in}}  &\sim e^{2i\omega r}r^{4iM\omega}\left(1 - \frac{i[16M^2\omega^2 - l(l+1)]}{2\omega r} + \ldots\right)
\end{align}
Using the relation \refeq{rstar} we can write
\beq
  e^{2i\omega r}r^{4iM\omega} = e^{2i\omega\rstar}\left[\frac{r}{C(r-2M)}\right]^{4iM\omega} 
    \sim \frac{e^{2i\omega\rstar}}{C^{4iM\omega}}\left[1 + \frac{8iM^2\omega}{r} + \ldots \right]
\eeq
The outgoing solution approaches a constant whereas the ingoing
solution, which is proportional to $e^{2i\omega\rstar}$, decays
exponentially for damped modes. The phase function has the leading
order behaviour
\beq
  g = \frac{\psi'}{\psi} \sim \frac{\frac{il(l+1)}{2\omega r^2}C_{\mrm{out}} 
      + 2i\omega e^{2i\omega \rstar}C_{\mrm{in}}}{C_{\mrm{out}} +  e^{2i\omega \rstar}C_{\mrm{in}}}
\eeq 
so that, for damped modes, 
\beq
  g \sim g_{\mrm{out}} \sim \frac{il(l+1)}{2\omega r^2}
\eeq
making manifest the problem of separating a general solution from a
purely outgoing one. Introducing the complex $r$ coordinate given in equation
\refeq{rcplx} we turn the damping exponentials into simple trigonometric
functions so that for large $\cpr$ and $C_{\mrm{in}}\neq 0$
\beq
  g \sim \frac{2i\omega e^{2i|\omega|\cpr}C_{\mrm{in}}}{C_{\mrm{out}} +  e^{2i|\omega|\cpr}C_{\mrm{in}}}
\eeq
For $C_{\mrm{in}} = 0$, \ie\ the outgoing solution, we have instead
\beq
  g \sim \frac{il(l+1)\omega}{2|\omega|^2\cpr^2} \rightarrow 0
\eeq
It is clear that if the boundary condition $g=0$ is imposed
``exactly at infinity'', there is no contamination of the unwanted
solution whatsoever. However, one may ask whether this preferred state
of affairs prevails if one is forced to impose the boundary conditions
at some large but finite radius $\cpr$. Putting the
expansions equal we find that the asymptotic condition for an outgoing
solution to coincide with a generic solution at some given radial
coordinate $\cpr=\cpr_{\infty}$ (say) is
\beq\
  e^{2i|\omega|\cpr_{\infty}}C_{\mrm{in}} \sim \frac{l(l+1)}{4|\omega|^2\cpr_{\infty}^2} C_{\mrm{out}}
\eeq
We see that for any finite $\cpr_{\infty}$ there exist complex
$C_{\mrm{in}}$ and $C_{\mrm{out}}$ such that numerical confusion
between the wanted $C_{\mrm{in}}=0$ solution and a mixed in- and
outgoing solution arise\footnote{By this we do not mean that there is
necessarily an exact coincidence of a mixed in/outgoing solution and
the purely outgoing solution, only that two such solutions are
numerically indistinguishable.}. However, it is also clear that this
solution has
$C_{\mrm{in}}\sim(|\omega|\cpr_{\infty})^{-2}C_{\mrm{out}}$ and is
therefore very close to the wanted solution. Thus, the only situation
where this can cause worries is when the ingoing solution is rapidly
decaying as a function of radius in some region (so that when
integrating from large radii this solution blows up). Such a blow up
will indeed happen in black hole spacetimes and one should also be
careful if stars more compact than $R\lesssim 3M$ are considered (due
to the peak in the effective potential).

\subsection{Numerical implementation}

\begin{figure}
  \includegraphics[width=0.40\textwidth,
  clip]{modes226_I.eps}\phantom{i}
  \includegraphics[width=0.62\textwidth, clip]{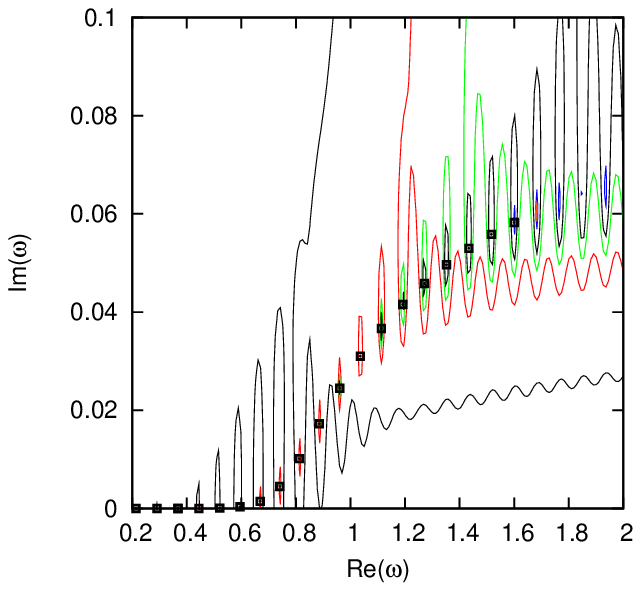}
  \caption{We plot the contours of
  $|g_s^{\mrm{tot}}(\tilde\omega)|$, where $\tilde\omega=\omega\sqrt{R^3/3M}$. 
  Quasi-normal modes correspond to zeros
  of this function and appear as ``islands'' in the plot.  In the
  left panel we show the region in the complex $\tilde\omega$-plane
  containing the interface modes. The shaded areas have not been
  covered due to convergence problems, mainly in the
  interior code, see main text for a discussion. Modes are displayed
  as black squares. The right panel shows the region containing the
  trapped modes and (a subset of) the curvature
  modes.}\label{fig:modes}
\end{figure}

Performing the compactification as described in the previous section, \ie\
changing the radial coordinate to
\beq\lbeq{xdef}
  x= -\frac{r-R-r_2}{r-R+r_2} \qquad \Leftrightarrow \qquad r = R + r_2\frac{1-x}{1+x}
\eeq
and using the phase function \refeq{phaseg} we obtain the equation
\beq
  g_{,x} = \frac{2r_2}{(x+1)^2}g^2 + e^{-2\nu}\frac{4r_2}{(x+1)^2}\left(\frac{M}{r^2}-i\omega\right)g
    + e^{-2\nu}\frac{2r_2}{(x+1)^2}\left(\frac{6M}{r^3}-\frac{l(l+1)}{r^2}\right)
\eeq
for the exterior vacuum perturbations.  This equation is solved by the
same routine that was applied to the toy problem of the previous
section, supplying a function $g^E_s(\omega) = g(x=1,\omega)$.  The
interior is governed by the wave equation \refeq{wbs} with the
potential given by \refeq{Upf}. Since our main aim here is to
demonstrate the usefulness of the approach to the exterior
perturbations we used a very simple code for the interior problem. The
equations were taken from \cite{ks:relastaxial} and are listed in the
appendix. We integrated the wave function in Schwarzschild coordinates
using an off the shelf Runge-Kutta routine. At the surface of the star
the phase function was evaluated, remembering eq.\ \refeq{dr} thus
giving rise to the function $g^I_s(\omega) = g(r=R, \omega)$. Now
$g^E_s(\omega)$ and $g^I_s(\omega)$ are guaranteed for every $\omega$
to satisfy the boundary conditions at infinity and $r=0$
repectively. It follows that the quasi-normal mode frequencies are
determined by the continuity of $g$ at the surface of the star, and
are hence given by the roots of 
\beq
  g_s^{\mrm{tot}} = g_s^E(\omega) - g_s^I(\omega) = 0
\eeq
We used the same strategy to find the modes as in the toy problem. In
figure \ref{fig:modes} we display contour plots of
$|g_s^{\mrm{tot}}|$. The modes appear as ``islands'' in this plot and
using the locations of these islands as initial guesses for our
M\"uller root finder we obtain the roots (\ie\ the quasi-normal mode
frequencies). A sample of the obtained frequencies are given in table
\ref{tab:freqs}. They are shown in figure \ref{fig:modes} as black squares. 
All results agree well with those of Kokkotas
\cite{kokkotas:axialmodes,kokkotas:axialmodescorr} 
and exactly with those of Tominaga, Saijo and Maeda~\cite{tsm:gw}. In
the left panel of figure
\ref{fig:modes} we show the interface modes. We have managed to locate
eight such modes. This raises the question if these modes constitute
an infinite family. For the most rapidly damped modes in this class we
encounter convergence problems both in the interior and the exterior
routines. The exterior problem is, however, less severe than the
interior one and a relative accuracy of about one part in $10^9$ in
the eigenfunction is achieved with moderate resolution. The interior
solution is more difficult to determine accurately. A quick glance at
the eigenfunctions reveal why, they are rapidly oscillating and
exponentially growing suggesting that a phase function approach would
be beneficial also for the interior problem. As stressed above, we are
mainly concerned with the methods here and did not consider it
relevant to try to improve the accuracy of the interior solutions for
this paper. From a theoretical point of view it would, however, be
interesting to settle the question on the number of interface modes.
In the right panel of figure \ref{fig:modes} we display the trapped
and curvature modes. For these modes we do not have any convergence
problems.

\ncd{\nlt}{\\[-5pt]}
\ncd{\trap}{$^\dagger$}
\ncd{\iface}{$^\ddagger$}
\begin{table}
\begin{center}
\begin{tabular}{|l l l | l l l|}
\hline
 & $\Re(\tilde\omega)$ & $\Im(\tilde\omega)$ & & $\Re(\tilde\omega)$ &
 $\Im(\tilde\omega)$ \\
\hline
\trap & $2.138639\E{-1}$ & $2.432061\E{-9}$ &  & $1.271223$        & $4.583803\E{-2}$ \nlt
\trap & $2.910115\E{-1}$ & $7.747252\E{-8}$ &  & $1.352193$        & $4.963654\E{-2}$ \nlt
\trap & $3.679986\E{-1}$ & $1.072543\E{-6}$   &  & $1.434156$        & $5.297615\E{-2}$ \nlt
\trap & $4.446349\E{-1}$ & $9.518605\E{-6}$   &  & $1.516856$        & $5.584165\E{-2}$ \nlt
\trap & $5.206390\E{-1}$ & $6.339316\E{-5}$   &  & $1.600034$        & $5.824413\E{-2}$ \nlt
 & $5.955793\E{-1}$ & $3.380602\E{-4}$  & \iface & $1.451160$        & $5.401143\E{-1}$ \nlt
 & $6.689607\E{-1}$ & $1.432267\E{-3}$  & \iface & $1.73434$         & $1.39451$ \nlt
 & $7.409095\E{-1}$ & $4.508010\E{-3}$  & \iface & $2.01107$         & $2.29027$ \nlt
 & $8.128027\E{-1}$ & $1.013574\E{-2}$  & \iface & $2.29394$         & $3.19756$ \nlt
 & $8.859040\E{-1}$ & $1.725926\E{-2}$  & \iface & $2.5798$          & $4.1068$ \nlt
 & $9.603731\E{-1}$ & $2.448020\E{-2}$  & \iface & $2.866$           & $5.016$ \nlt
 & $1.036081$       & $3.102749\E{-2}$  & \iface & $3.153$           & $5.924$ \nlt
 & $1.113082$       & $3.668522\E{-2}$  & \iface & $3.448$           & $6.846$ \nlt
 & $1.191470$       & $4.155932\E{-2}$  &  & & \\
\hline
\end{tabular}
\end{center}
\caption{Dimensionless frequencies $\tilde\omega=\omega\sqrt{R^3/3M}$ of the axial $l=2$ modes 
  in a $R=2.26M$ constant density star. The trapped modes, defined to
  be those whose frequency satisfy $\Re(\omega)^2<V_{\mrm{max}}$, are
  marked with a \trap. Here $V_{\mrm{max}} \approx 0.1513/M^2$ is
  defined to be the value of the effective potential in the
  Regge-Wheeler equation at the peak. The interface modes are marked
  by \iface. All digits shown should be correct and in many cases the
  accuracy is much better than displayed.}
\label{tab:freqs}
\end{table}

\section{Conclusions}

We have discussed a new approach to the quasinormal-mode problem in
general relativity. By combining a characteristic formulation of the
perturbation equations with the integration of a suitable
phase-function for a complex valued radial coordinate, we have
reformulated the standard outgoing-wave boundary condition as a zero
Dirichlet condition. This brings a number of important advantages over
previous strategies. The characteristic formulation permits coordinate
compactification, which means that we can impose the boundary
condition at future null infinity. The phase function avoids
oscillatory behaviour in the solution, while the use of a complex
radial variable allows a clean distinction between out- and ingoing
waves. We have demonstrated that the method is straightforward to
implement, and our analysis of two simple toy problems shows that it
can lead to high precision numerical results. It is worth noting that
the generalisation to unstable modes is straightforward, only
requiring an alteration of the integration contour.

Even though the numerical results we have presented are in themselves
of no great interest, the new method may represent a breakthrough in
this area. Most importantly, we have every reason to believe that it
should generalise to the problem of rapidly rotating relativistic
stars. In that case one would no longer deal with a simple
one-dimensional wave equation in the exterior vacuum. The lack of
workable implementations of the required outgoing-wave conditions in
that problem has been holding back progress for many years. Our new
approach may be the key that unlocks this problem. Of course, we are
still some steps away from implementing our new ideas for rotating
star spacetimes. Most importantly, we need to consider how the
different ingredients in our prescription generalise to this more
complicated problem.  This is an interesting challenge and we would
hope to make progress on it in the near future.

\section*{Acknowledgements}
LS gratefully acknowledge support by a Marie Curie Intra-European
Fellowship, contract number MEIF-CT-2005-009366. This work was also
supported by PPARC through grant numbers PPA/G/S/2002/00038 and
PP/E001025/1.  NA acknowledges support from PPARC via Senior Research
Fellowship no PP/C505791/1. Finally, we acknowledge support from the
EU-network ILIAS providing opportunity for valuable discussions with
our European colleagues.

\appendix
\section{Constant density sphere -- $w$-modes.}

We test the code by computing the $w$-modes of a uniform sphere of
mass $M\neq 0$. In the interior the background line element can be written 
\beq
  \d s^2 = -e^{2\nu}\d t^2 + e^{2\lambda}\d r^2 + r^2(\d\theta^2 + \sin^2\theta\d\phi)
\eeq
where
\begin{align}
  e^\nu &= \frac12\left(3\sqrt{1-\frac{2M}{R}} - \sqrt{1-\frac{2Mr^2}{R^3}}\right) \\
  e^{-\lambda} &= \sqrt{1-\frac{2M r^2}{R^3}} 
\end{align}
see \eg\ \cite{schutz:gr}. The perturbation equations can be put in the form \cite{ks:relastaxial}
\ncd{\Xcal}{\mathcal{X}}
\begin{align}
  r\frac{\d\Xcal_1}{\d r} &= -(l+2)\Xcal_1 - e^{\lambda-\nu+\nu_c}\Xcal_2 \lbeq{X1} \\
  r\frac{\d\Xcal_2}{\d r} &= -e^{\lambda-\nu-\nu_c}\left[(l-1)(l+2)e^{2\nu} - \omega^2r^2\right]\Xcal_1 - (l-1)\Xcal_2 \lbeq{X2}
\end{align}
where
\begin{align}
\Xcal_1 = i\omega r^{-(l+1)}\psi \\
\Xcal_2 = -e^{\nu_c-\nu}\omega^2r^{-l}\varphi 
\end{align}
and $\nu_c$ is the central value of $\nu$. The function $\varphi$ can
in this context be viewed as auxiliary and defined by equation \refeq{X1}

At the centre of the star the regular solution behave as
\begin{align}
  \Xcal_1 &= \hat\Xcal_1\left\{1 - \frac{e^{-2\nu_c}\omega^2 + (l+2)\tsfrac{M}{R^3}(e^{-\nu_c} - 2l)}{2(2l+3)}r^2  
        + O(r^4)\right\} \lbeq{exp1}\\
  \Xcal_2 &= \hat\Xcal_1\left\{-(l+2) + \frac{(l+4)e^{2\nu_c}\omega^2 - (l+2)(l-1)\tsfrac{M}{R^3}(e^{-\nu_c}+2l+6)}{2(2l+3)}r^2
        + O(r^4)\right\} \lbeq{exp4} 
\end{align}
where $\hat\Xcal_1$ is an arbitrary constant.



\begin{thebibliography}{10}

\bibitem{chandra:cfs}
S. {Chandrasekhar}, Physical Review Letters {\bf 24},  611  (1970).

\bibitem{fs:cfs}
J.~L. {Friedman} and B.~F. {Schutz}, \apj {\bf 222},  281  (1978).

\bibitem{ks:qnm}
K.~D. Kokkotas and B.~G. Schmidt, \lr {\bf 2},  2  (1999),
  \eprint{qr-qc/9909058}.

\bibitem{andersson:bhqnm}
N. {Andersson}, Royal Society of London Proceedings Series A {\bf 439},  47
  (1992).

\bibitem{aks:qnm}
N. {Andersson}, K.~D. {Kokkotas}, and B.~F. {Schutz}, \mnras {\bf 274},  1039
  (1995), \eprint{gr-qc/9503014}.

\bibitem{bwmb:rot}
E. {Berti}, F. {White}, A. {Maniopoulou}, and M. {Bruni}, \mnras {\bf 358},
  923  (2005).

\bibitem{sf:neutral}
N. {Stergioulas} and J.~L. {Friedman}, \apj {\bf 492},  301  (1998),
  \eprint{gr-qc/9705056}.

\bibitem{msb:neutral}
S.~M. {Morsink}, N. {Stergioulas}, and S.~R. {Blattnig}, \apj {\bf 510},  854
  (1999), \eprint{gr-qc/9806008}.

\bibitem{ye:rotcowling}
S. {Yoshida} and Y. {Eriguchi}, \apj {\bf 515},  414  (1999),
  \eprint{astro-ph/9807254}.

\bibitem{mina:thesis}
A. Maniopoulou, Ph.D. thesis, School of {M}athematics, {U}niversity of
  {S}outhampton, 2005.

\bibitem{ga:quickndirty}
K. {Glampedakis} and N. {Andersson}, Classical and Quantum Gravity {\bf 20},
  3441  (2003), \eprint{gr-qc/0304030}.

\bibitem{ccm:V}
M.~R. {Dubal}, R.~A. {{d}'{I}nverno}, and J.~A. {Vickers}, \prd {\bf 58},
  044019  (1998).

\bibitem{kalle:phase}
K.-E. Thylwe, \jpa {\bf 38},  10007  (2005).

\bibitem{cf:osc}
S. Chandrasekhar and V. Ferrari, \prsla {\bf 432},  247  (1991).

\bibitem{kokkotas:axialmodes}
K.~D. {Kokkotas}, \mnras {\bf 268},  1015  (1994).

\bibitem{bbm:characteristic}
H. {Bondi}, M.~G.~J. {van der Burg}, and A.~W.~K. {Metzner}, Royal Society of
  London Proceedings Series A {\bf 269},  21  (1962).

\bibitem{sachs:characteristic}
R.~K. {Sachs}, Royal Society of London Proceedings Series A {\bf 270},  103
  (1962).

\bibitem{boyd:spectral}
J.~P. Boyd, {\em Chebychev and {F}ourier Spectral Methods}, second revised ed.
  (Dover Publications, New York, USA, 2001).

\bibitem{cf:gwresonance}
S. Chandrasekhar and V. Ferrari, \prsla {\bf 434},  449  (1991).

\bibitem{akk:oscillation}
N. Andersson, Y. Kojima, and K.~D. Kokkotas, \apj {\bf 462},  855  (1996).

\bibitem{tsm:gw}
K. {Tominaga}, M. {Saijo}, and K.-I. {Maeda}, \prd {\bf 60},  024004  (1999),
  \eprint{gr-qc/9901040}.

\bibitem{ks:w-modes}
K.~D. Kokkotas and B.~F. Schutz, \mnras {\bf 255},  119  (1992).

\bibitem{lns:w2}
M. {Leins}, H.-P. {Nollert}, and M.~H. {Soffel}, \prd {\bf 48},  3467  (1993).

\bibitem{ks:relastaxial}
M. Karlovini and L. Samuelsson, \cqg {\bf 24},  3171  (2007),
  \eprint{gr-qc/0703001}.

\bibitem{kokkotas:axialmodescorr}
K.~D. {Kokkotas}, \mnras {\bf 277},  1599  (1995).

\bibitem{schutz:gr}
B.~F. Schutz, {\em A first course in general relativity} (Cambridge University
  Press, Cambridge, U.K., 1985).

\end{thebibliography}

\end{document}